\documentclass{article}
\usepackage{authblk}

 \usepackage{amsmath}
\usepackage{amsfonts}

\usepackage{breqn} 
\usepackage[english]{babel}
\usepackage{capt-of}
\usepackage[noadjust]{cite}
\usepackage{colortbl}
\usepackage{csquotes}

\usepackage{float}
\usepackage{graphicx}
\usepackage[margin=1in]{geometry}
\usepackage{multicol}
\usepackage{multirow}
\usepackage[utf8]{inputenc}

\usepackage[colorinlistoftodos,prependcaption]{todonotes}
\usepackage[normalem]{ulem}

\usepackage{url}

\usepackage{etoolbox}

\patchcmd{\abstract}{Abstract}{Executive Summary}{}{}

\definecolor{red}{rgb}{1.00,0.00,0.00}

\definecolor{blue}{rgb}{0,0.00,50}

\definecolor{purple}{rgb}{0,0.00,0}

\newcommand{\mbw}{{\mathbf{w}}}

\usepackage{blindtext}
\usepackage{scrextend}

\author[1]{Dagoberto Pulido}
\author[1]{Daniela Basurto}
\author[1]{Mayra Cándido}
\author[1,2]{Joaquín Salas\thanks{Corresponding Author: Joaquín Salas. CICATA Querétaro, Instituto Politécnico Nacional. Cerro Blanco 141, Colinas del Cimatario, Querétaro, CP 76090, México. {\tt jsalasr@ipn.mx}}}

\affil[1]{Instituto Politécnico Nacional, México}

\date{}

\begin{document}

\title{Geospatial Spread of the COVID-19 Pandemic in Mexico}

\maketitle

\begin{abstract}

COVID-19 is an infectious respiratory disease that the World Health Organization has declared a pandemic. Although a global phenomenon, there is the need to react locally with swift and informed actions, as some of the essential plans are highly dependent on factors such as culture, geography, laws, and customs. This paper presents our approach to mapping the geospatial spread of COVID-19 in Mexico at the state and municipal level, the highest allowed possible resolution in the publicly available dataset provided by the Health Ministry. To visualize the magnitude of the infection, we offer a map for the confirmed positive, pending, and deceases cases, while to support mobility, we provide a geospatial visualization of $R_t$, the basic reproduction number. This document describes the structure of the dataset, the software tools employed, and a description of the functionality of the maps, which definition we make publicly available.
\end{abstract}


\section{Introduction}

COVID-19 in an infectious respiratory disease that has been declared a pandemic by the United Nations (UN) World Health Organization (WHO)~\cite{cucinotta2020declares}. Affecting the lungs, it makes them stiff, thus difficulting breathing. Usually, the air we pull through the trachea branches into each lung. Then, it keeps splitting subsequently into smaller pipes, eventually reaching the bronchioles and afterward the alveoli, which are microscopic sacs of air~\cite{aguiar2020inside}. 
SARS-CoV-2 (the Severe Acute Respiratory Syndrome Coronavirus-2), the virus causing COVID-19, promotes the creation of an exudate that fills the alveoli impeding gas exchange~\cite{tian2020pathological}. In terms of the  COVID-19 infection, it is still too early to assess the portion of the population that is immune, is asymptomatic, requires hospitalization,   needs a ventilator, and eventually  perish~\cite{gonzalez2020sars}.
For the sake of discussion, consider the case of symptomatic people.
The most frequent symptoms at the beginning are fever, cough, and fatigue. Other signs that appear afterward include sputum production, myalgia, headache, hemoptysis, diarrhea, dyspnea, and lymphopenia~\cite{rothan2020epidemiology}.
Nonetheless, it is still unclear what circumstances cause a particular person to show symptoms, and significant statistics are overdue to provide some light.  Consider, for instance, that while a study suggests that up to 50\% of the population of Iceland is asymptomatic~\cite{gudbjartsson2020spread},
in India, a survey showed that 47.1\% of the infected population is asymptomatic and 35.3\% showed mild symptoms~\cite{thangaraj2020cluster}. In the meantime,  in a Boston shelter out of 408 people, 147 tested positive, and 361 reported no symptoms~\cite{baggett2020prevalence}.
Thus, to reign over confusion, we need to generate and spread trustable information quickly, reliably, and efficiently. 





Maps are a suitable solution as one may obtain a vast amount of information quickly~\cite{knudsen2020evolution}. There have been numerous attempts to develop visualizations of the spread of the COVID-19. For instance, Dong {\it et al.}~\cite{dong2020interactive}, at John Hopkins University, created a worldwide map of cases, including every country where there was at least one case.  To understand the complexity of such a task, one may want to consider that it includes the problems associated with gathering a vast amount of information from several centers of production, each with its data format and produced at different speeds~\cite{roser2020coronavirus}. Across the visualizations of the spread of COVID-19, one finds tradeoffs between the scope of the coverage and the level of detail provided. Although a milestone achievement, until recently, Mexico was represented in this and other worldwide maps as a single point.
Thus, there is a clear need to develop local and updated versions of maps to support the availability of reliable information using efficient and comprehensive coverage means. 












In this document, we describe maps to assess the magnitude of the epidemy and to guide the process of returning to normality. In the former case, we emphasize that keeping up with the spread of COVID-19 is a complex problem by itself. Just consider that Mexico has a population of 130 million people, which extends over 1.973 million km$^2$. Its geography includes 32 states, first-level administrative divisions in the country. In turn, each state is divided politically into more than 2450 municipalities, second-level administrative divisions. As the pandemic progress, governments have implemented lockdown actions to augment social distancing.
In consequence, the economic factors start increasing its relative importance. Just a result of the pandemic, the European Commission has projected a shrink in the EU GDP  of 7.4\%~\cite{kinross2020rapidly}, the US economy is in its way to reaching a 20\% of unemployment~\cite{atkeson2020will}, and the Chinese economy shrank 6.8\% in the first quarter of 2020, ending a 50 years long run of growth~\cite{khan2020china}. With these three engines of the global economy turned off, others will falter. So, there is an apparent need to open as soon as possible the economy while making sure that the health aspects are kept at bay. Thus, this work contributes to addressing the urgent need for rapid, updated, and reliable information at the municipality and state level.




\begin{figure*}
\centering
\begin{tabular}{cc}
\includegraphics[width=3in]{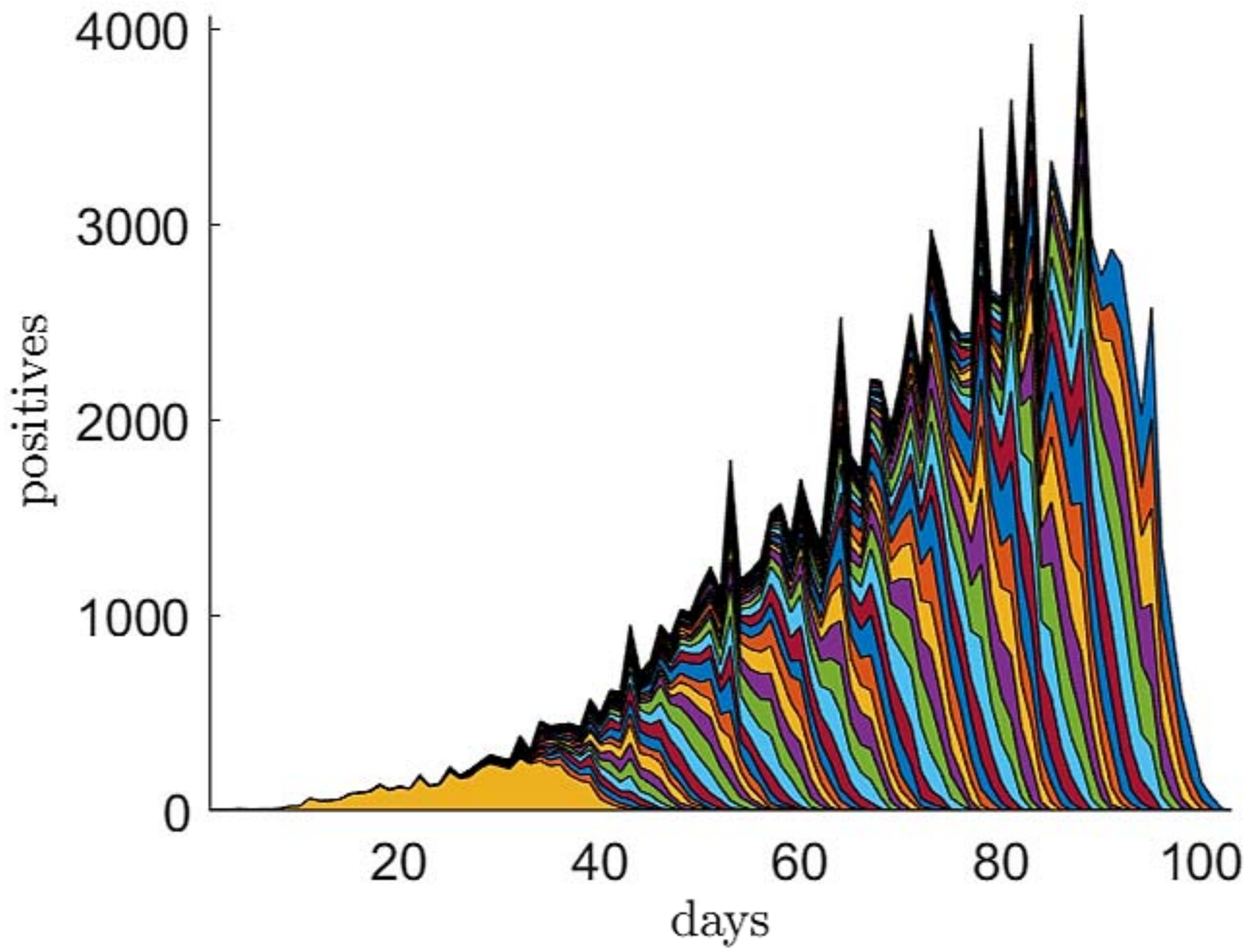}  & \includegraphics[width=3in]{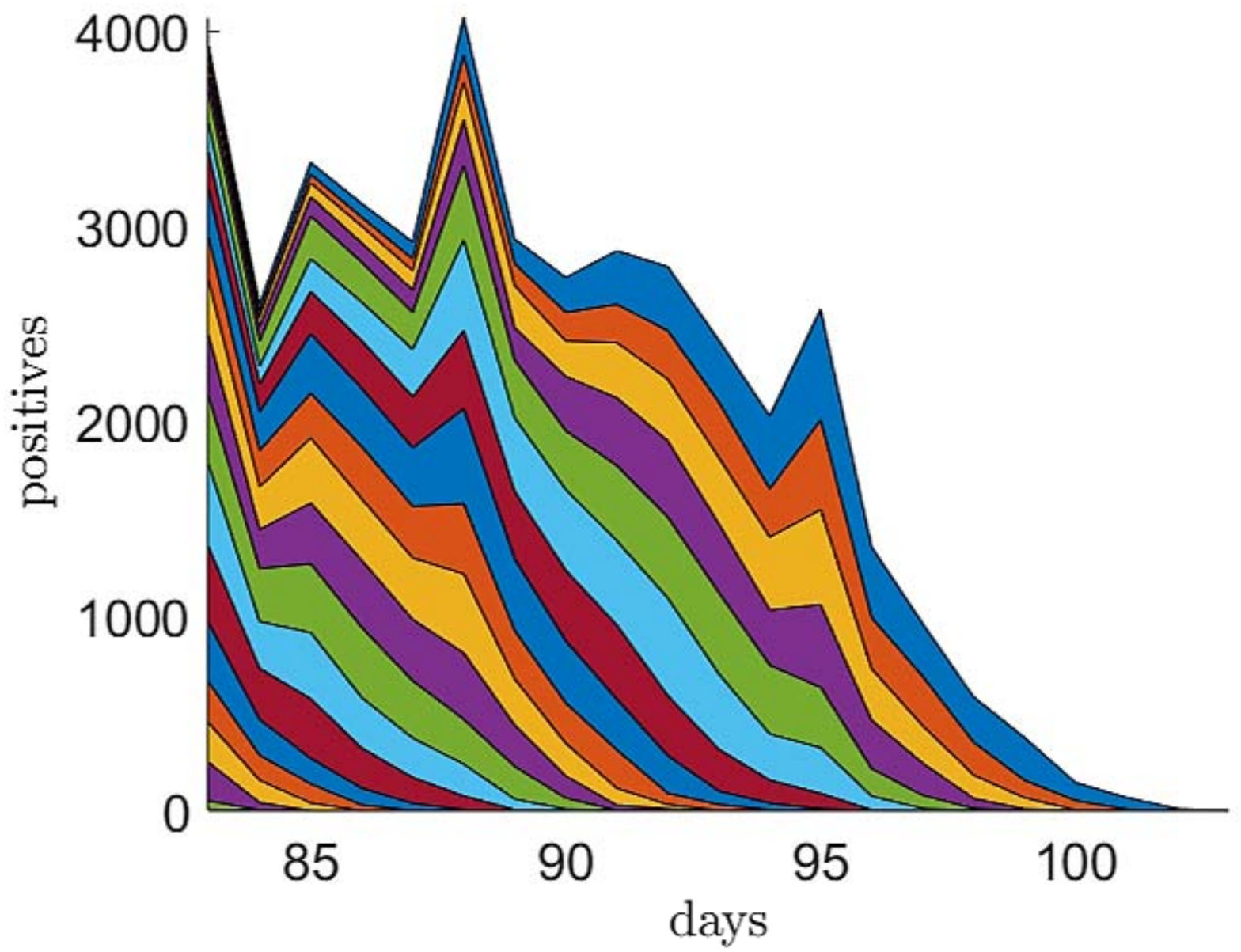} \\
(a) Full observation range & 
(b) Last 20 days\\
\end{tabular}
\caption{Delays in Acquiring the Data. Due to delays in diagnosis, testing,
and information processing flow, there is a delay of a considerable number of days between the date of first symptoms and availability for analysis.   
}
\label{fig:delays}
\end{figure*}

\section{Confirmed Positive, Pending and Deceased}

\begin{figure*}[t]
    \centering
    \begin{tabular}{c}
    \includegraphics[width=4.7in, trim=10 10 10 55, clip]{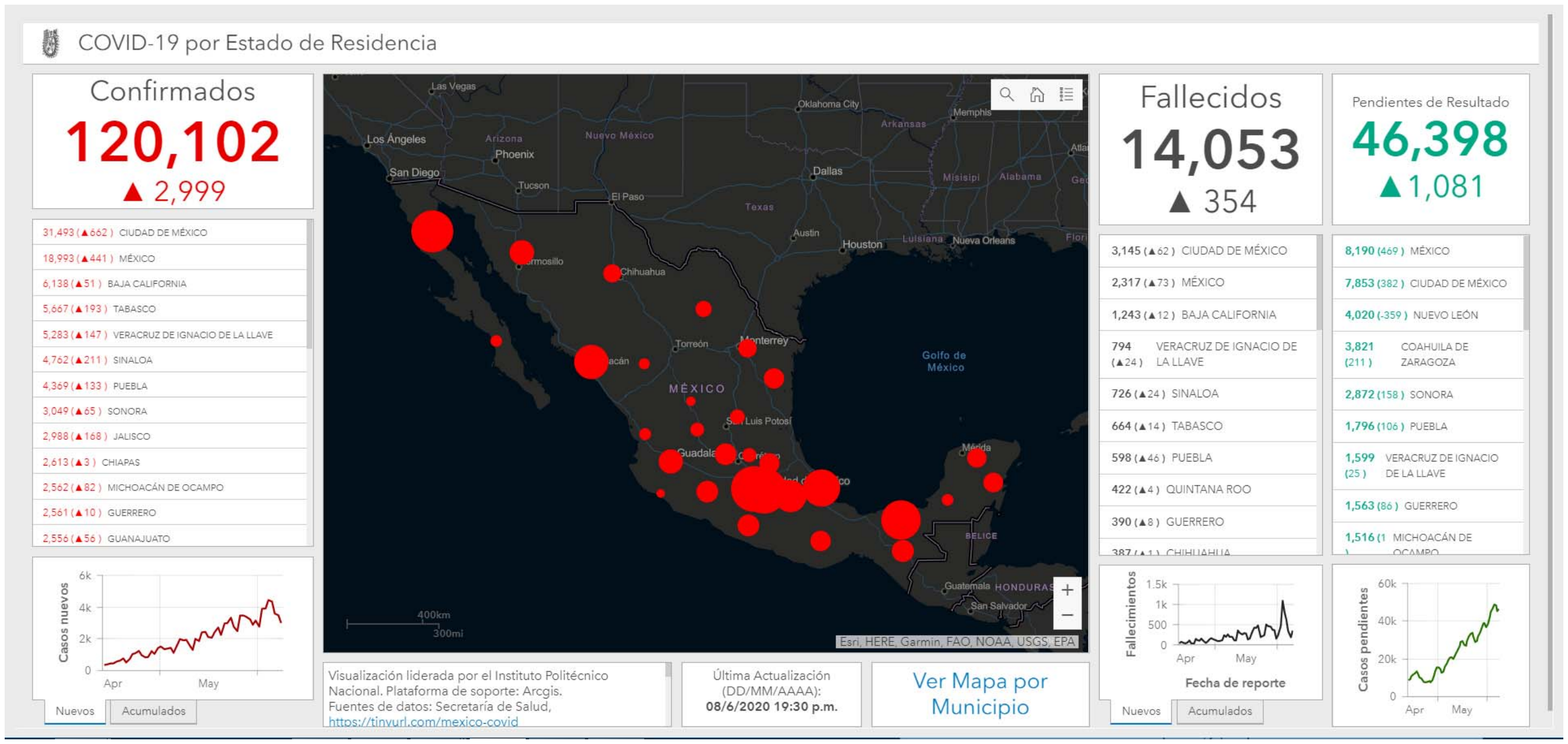} \\ (a)\\ \includegraphics[width=5in, trim=0 200 0 0, clip]{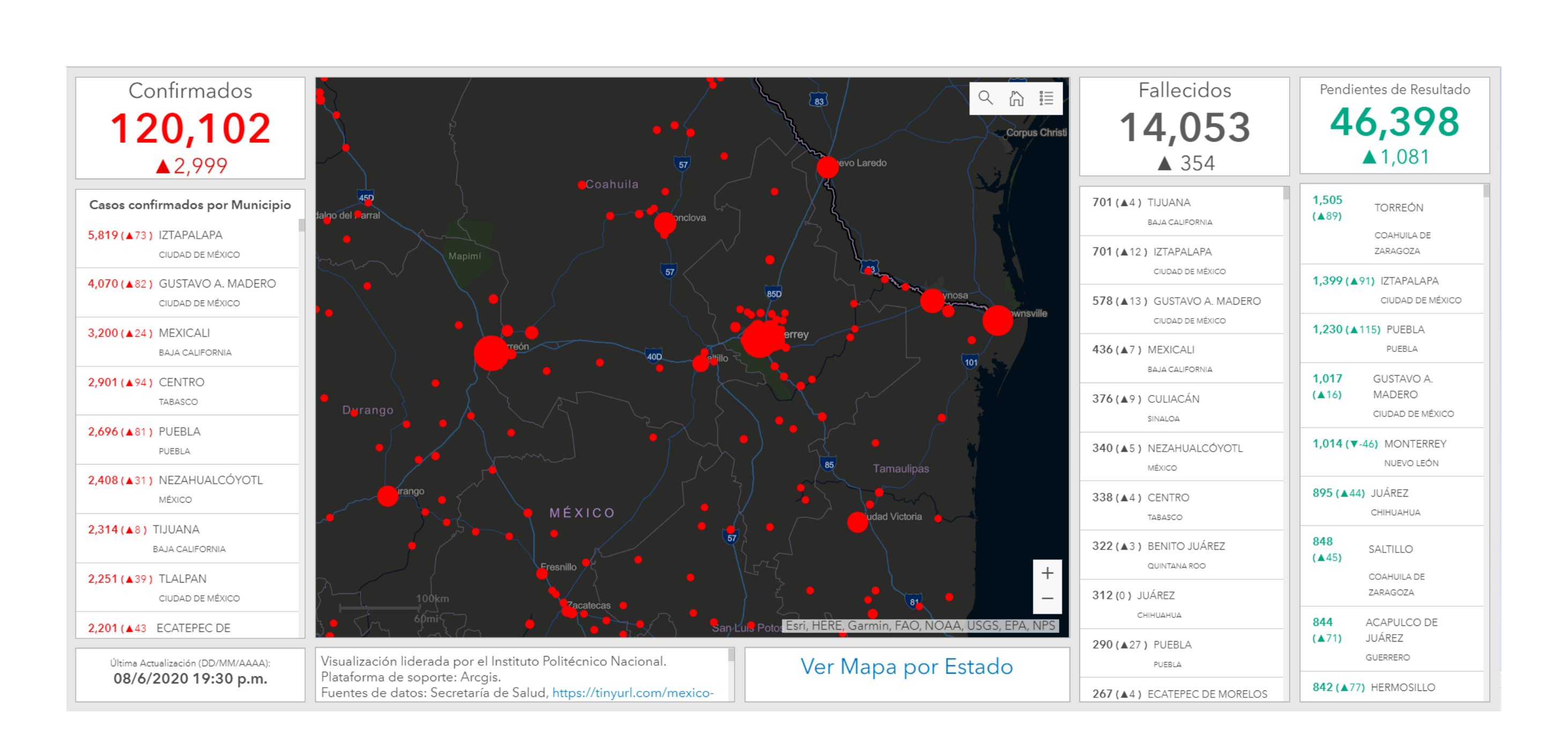}\\
    (b) \\
    \end{tabular}
    \caption{Confirmed, deceased and pending cases of COVID-19 in Mexico at state (a) and municipal(b) level.  }
    \label{fig:covid-spread-cases}
\end{figure*}

In this map, we show, at Municipio and State levels of granularity, the number of confirmed positive, pending, and deceased cases of COVID-19 and associate them with their place of residence  (see Figure~\ref{fig:covid-spread-cases}).

\subsection{Preprocessing the Data}
The Mexican government, through its Health Ministry, releases every day, since April 12, 2020, a dataset containing the registers of individual cases~\cite{dataset}. The dataset, which fields we describe in detail in Appendix~\ref{app:data}, is the source of information we employ in our maps. For our purposes, we extract patients' data, including their State of residence, their Municipio of residence, the result of their tests (confirmed, positive or negative, or pending) and date of death (when appropriated). Please note that, in the case of COVID-19, the generation of maps is a dynamic process that includes a large number of stages. Consider, for instance, the example of a patient that starts feeling COVID-19 symptoms. Once the patient decides to visit the care provider, a physician may determine a need for a test. Depending on the available infrastructure, the test may have to travel to a nearby location. After a few days, the results come back to the physician, who fills out a local agency report. The national network of health care providers collects these results and finally makes them public. In Figure~\ref{fig:delays}, we illustrate the daily update of positive confirmed cases occurring in the country. Note that there may be a delay of more than 20 days between a self-reported start of symptoms until the data is finally officially registered. These delays may make it harder to assess the magnitude of the epidemy and to develop plans of action.

To efficiently organize the information int the dataset, we create data tables employing FileMaker~\cite{filemaker}, a database management system that facilitates the incorporation of a  Graphical User Interface (GUI). The database consists of six tables: {\tt current}, {\tt previous}, {\tt municipalities}, {\tt states}, {\tt statistics.mu\-ni\-ci\-palities}  and {\tt statistics.states}. The {\tt current} and  {\tt previous} tables have a structure corresponding to the fields in the data file published by the government (see Appendix~\ref{app:data}). The {\tt municipalities} and {\tt states} tables contain the name, the latitude, and the longitude of the place to locate it on the map. The tables {\tt statistics.mu\-ni\-ci\-palities} and {\tt statistics.states} hold the number of positively confirmed, deaths,  pending results, difference of each of the indicators to the previous day, and boolean fields with indicators changing depending on whether the current value is greater or smaller to the day before. We calculate summaries by State and Municipio searching on the {\tt statistics.states} and  {\tt statistics.mu\-ni\-ci\-palities} tables.

\subsection{Data Visualization }
 We display the data as circles which size is proportional to the population they represent,  locating the geographic coordinates from ubicated points on the interactive maps in unities of latitude and longitude (see Figure \ref{fig:covid-spread-cases}). We also include timeline graphs alongside the state-level map (see Figure \ref{fig:covid-spread-cases}(a)). These graphs show chronologically the new and accumulated confirmed cases, daily and accumulated deaths, and cases with pending results since the Health Ministry released the public database. The user can select a state to filter its corresponding data within the map view, and this shows a state-specific graph. When there is no selection, the data for every state is added, showing a nation-wide timeline graph. 
It is important to note that the graphs for confirmed cases show the data on the dates reported by the Health Ministry. This date may or may not be the date when the care provider registered the test result, as the dataset does not include that information. In the case of deaths, the official dataset usually reports them days after the passing occurred, but provide the date of death; thus, we generate a timeline graph by the actual date. Given this delay in reporting, the graphs for deaths are prone to changes with each daily update, and numbers for recent dates are most likely to increase.


During the updating process, we compare our results with the official report as an initial quality check. Afterward, we use the following policies: a) When a record does not have a Municipio of residence, we assign it to the State capital; b) if there is a missing State of residence information, the record is marked as unidentified and assigned to the country capital. We calculate daily the difference between the confirmed positives, pending, and deaths with the previous day. To improve the visualization impact, we include an arrow upward $\Uparrow$ whenever there is an increase; on the contrary, we show a downward arrow  $\Downarrow$ to signal a decrement in the indicator. After processing the data, we export the corresponding data file and upload it to overwrite the data layer in the map project. We register the date and hour of the update.  The map shows the municipalities with at least one case of COVID-19. As well as an ordered list, from high to low, of the incidence of the disease. 

We separately process the data we use for the State-level graphs. Even when the public dataset is updated daily, the records within it lack the date when the Health Minister first included them. Therefore, we collect all the uploaded datasets to compare each with the previous one and determine which records are new. We do this to obtain the daily new and total positive cases and cases awaiting the state's results. In the case of deaths, we only use the last updated public dataset as the date of death is included in each record. We process this information in {\bf R} (the computer language) and output the collected data as CSV (comma separated variables) files that specify the date and state for each statistic. We then upload these CSV files to the map project to generate the graphs. 
\section{Basic Number of Reproduction}


In the previous section, we addressed the problems of increasing awareness on the magnitude of the epidemy in terms of the number of people confirmed positive, suspected, and deceased at the municipal and state level. Rapid access to that information supports the mobilization of sanitary and psychological resources. A complementary piece of information to guide the public on the magnitude of the epidemic is the basic reproduction number, $R_t$, which represents the indicator that should govern society on whether the contagious increase, $R_t>1 $, or decrease, $R_t <1$.

\begin{figure}[h]
    \centering
    \begin{tabular}{ccc}
    \begin{minipage}{1in}
    \includegraphics[width=2in,trim=0 35 90 0, clip]{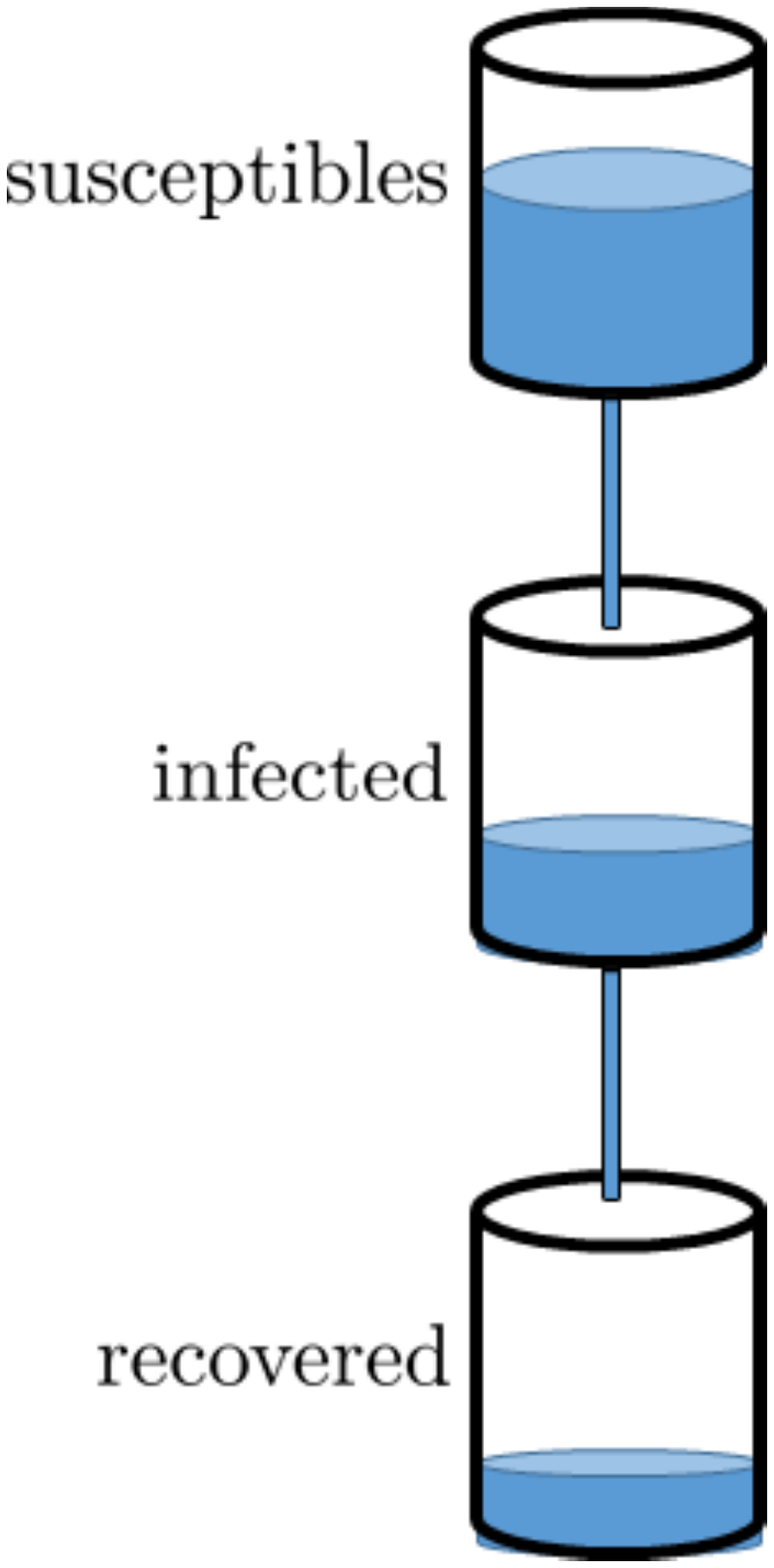}
    \end{minipage}&
    \begin{minipage}{2.5in}
          \includegraphics[width=2.5in]{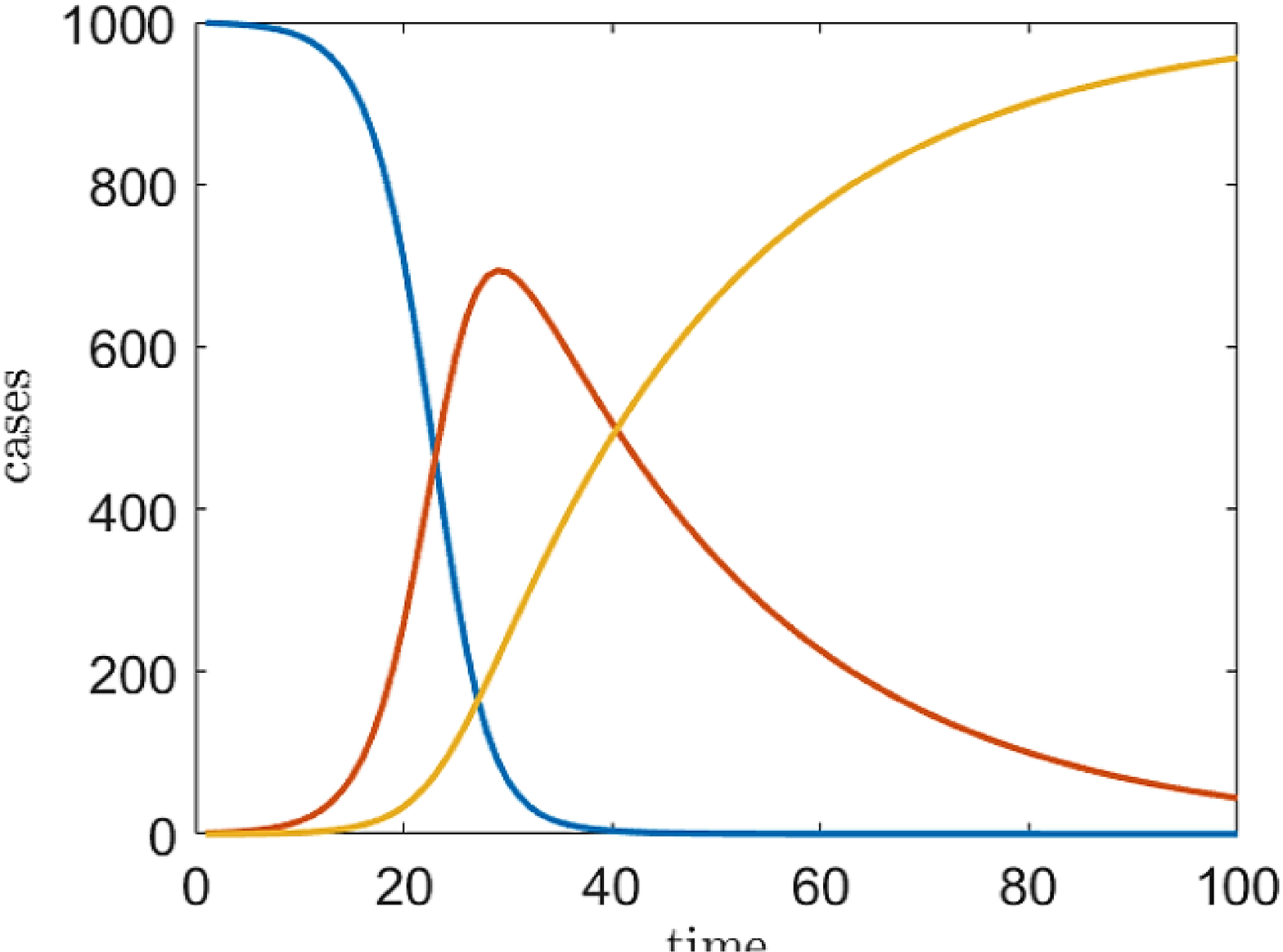}
          \end{minipage}&
          \begin{minipage}{2.5in}
          \includegraphics[width=2.5in]{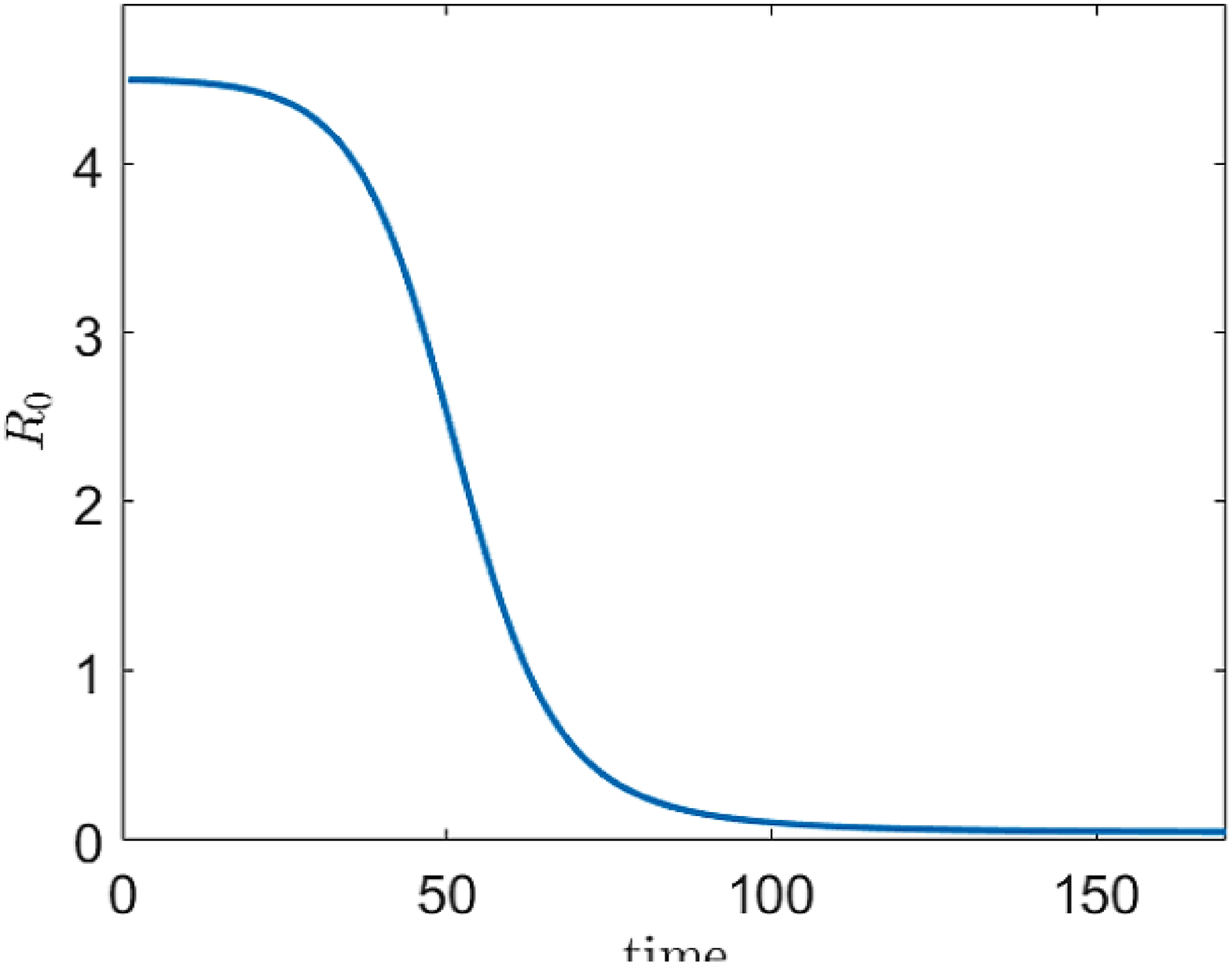}
          \end{minipage}\\
          (a) & (b) & (c) \\
        \end{tabular}
    \caption{SIR Model. In SIR, one divides the population into susceptibles, infected and recovered, and may move those categories subsequently (a). The variation in the population among the groups obey differential equations (b), commanded by the value of $R_t$ (c).}
    \label{fig:SIR}
\end{figure}

\subsection{Interpretation of $R_t$ in the SIR Model}
Back in 1927, Kermack and McKendrick~\cite{kermack1927contribution} proposed the SIR model to interpret infectious diseases. They assumed that the whole number of individuals $N$ in a population could be divided disjointly in suspected $S$, infected $I$, and recovered $R$. Another assumption in the model is that susceptibles may become infected, infected may become recovered, and recovered do not become infected again (see Figure ~\ref{fig:SIR}). 


This setting leads to a dynamic system where the number of susceptibles decreases over time as a fraction $a$ of the number of interactions between susceptibles and infected, where we are dropping the dependency on $t$ for clarity, as
\begin{equation}
\partial S/\partial t
= - a \frac {I S} {N}.
\end{equation}
Meanwhile, the number of infected increases by the interaction
between infected and susceptibles but decreases by the fraction of infected people recovering as
\begin{equation}
\partial I/\partial t
= a \frac {I S} {N} - b I.
\label{eq:dinfected}
\end{equation}
Finally, the number of recoveries increases as a proportion of the
infected as
\begin {equation}
\partial R/\partial t
= b I.
\end {equation}
Here $a$ and $b$ are proportionality constants reflecting the inverse of the number of interactions in units of time. Their units are {\it people}$^{-1}\cdot$ {\it time}$^{- 1}$.

One may express (\ref{eq:dinfected}), the equation expressing the rate at which  the number of infected changes,  as
\begin{equation}
\partial S/\partial t
= I \left (a \frac {S} {N} - b \right),
\end{equation}
which has a solution
\begin{equation}
I = I_0 \displaystyle {e^{\left (a \frac {S} {N} - 
b \right) t}},
\end{equation}
where $I_0$ is the initial number of infected, 
that grows indefinitely if the exponent is positive and  
 fades if it is negative.
One may express the stability condition as
\begin{equation}
\begin{array} {ccc}
a \frac{S} {N} - b = 0 & \mbox {ó} & \frac {aS} {bN} = 1.
\end {array}
\end{equation}
Thus, one may express the basic reproduction number 
or average number of secondary cases caused by an infected person, as
\begin{equation}
R_t = \frac {aS} {b N},
\end{equation}
which indicates that the number of infected will increase, when $R_t>1$, or decrease, when $R_t <1$.

\subsection{Bayesian Solution for $R_t$}
To estimate the most likely value for a random variable $R_t$, one would have to have some kind of  tracking capability to solve among other factors who is infected, when the infection started, what people contacted the infectee, and who among these people contracted the infection. In addition, there would be the need follow this process during the evolution of the epidemy, something that may be harder as the infection spread. 
Thus, one should make some general assumptions to model the phenomenon in practice due to these difficulties. One of them is that the rate at which an infected person spread the disease to others follow a Poisson process described as~\cite{cori2013new}
\begin{equation}
P(I_t\mid I_0, \dots, I_t, \mbw, R_t) =  \frac{(R_t \Lambda_t)^{I_t} e^{-R_t \Lambda_t}}{I_t!},
\end{equation}
where $I_k$ corresponds to the number of infections at time $k$,  $\mbw$ defines the serial interval for the spread, in the form of a mass probability density, and one defines the expected number of infectious as~\cite{cori2013new}
\begin{equation}
\mathbb{E} [I_t] = R_t \sum_{s=1}^t I_{t-s} w_s = R_t \Lambda_s. 
\end{equation}
There are several approaches to estimate the parameters that maximize the probability distribution, including maximum likelihood, maximum a posteriori probability, and a general Bayesian approach~\cite{prince2012computer}. The latter is interesting as it permits us to present the results as a general probability distribution without committing to a point estimate. To that end, one needs to propose a {\it prior}, which form one has to assume known. Cori {\it et al.}~\cite{cori2013new} propose to use a Gamma distribution for $P(R_{t, \tau})$ as 
\begin{equation}
P(R_{t, \tau}) = \frac{R_{t,\tau}^{a-1}}{\Gamma{(a)} b^a}  e^{-R_{t,\tau}/b},
\end{equation}
where $a$ and $b$ correspond respectively to the shape and scale. This arrangement is fortunate as the Poisson distribution along with a Gamma distribution as its conjugate results in a Gamma distribution expressed as~\cite{cori2013new} 
\begin{dmath}
P(I_{t - \tau +1}, \dots, I_t, R_{t,\tau}\mid I_0, \dots, I_{t-\tau}, \mbw) 
= R_{t,\tau}^{\alpha - 1}
    e^{R_{t,\tau}/ \beta} 
\prod_{s = t - \tau + 1}^t \frac{\Lambda_s^{I_s}}{I_s!} \frac{1}{\Gamma(a)b^a}.
\end{dmath}
From where one may obtain  the shape and scale  parameters, which transformed into the mean and standard deviation results in
\begin{equation}
\alpha = a + \displaystyle{\sum_{s=t - \tau +1} I_s} \mbox{ y }
\beta = \frac{1}{\displaystyle{\sum_{s=t-\tau+1}^{t} \Lambda_s} +\frac{1}{b}}.
\end{equation}

\begin{figure}[t]
    \centering
    \includegraphics[width=4in,trim=0 80 80 0, clip]{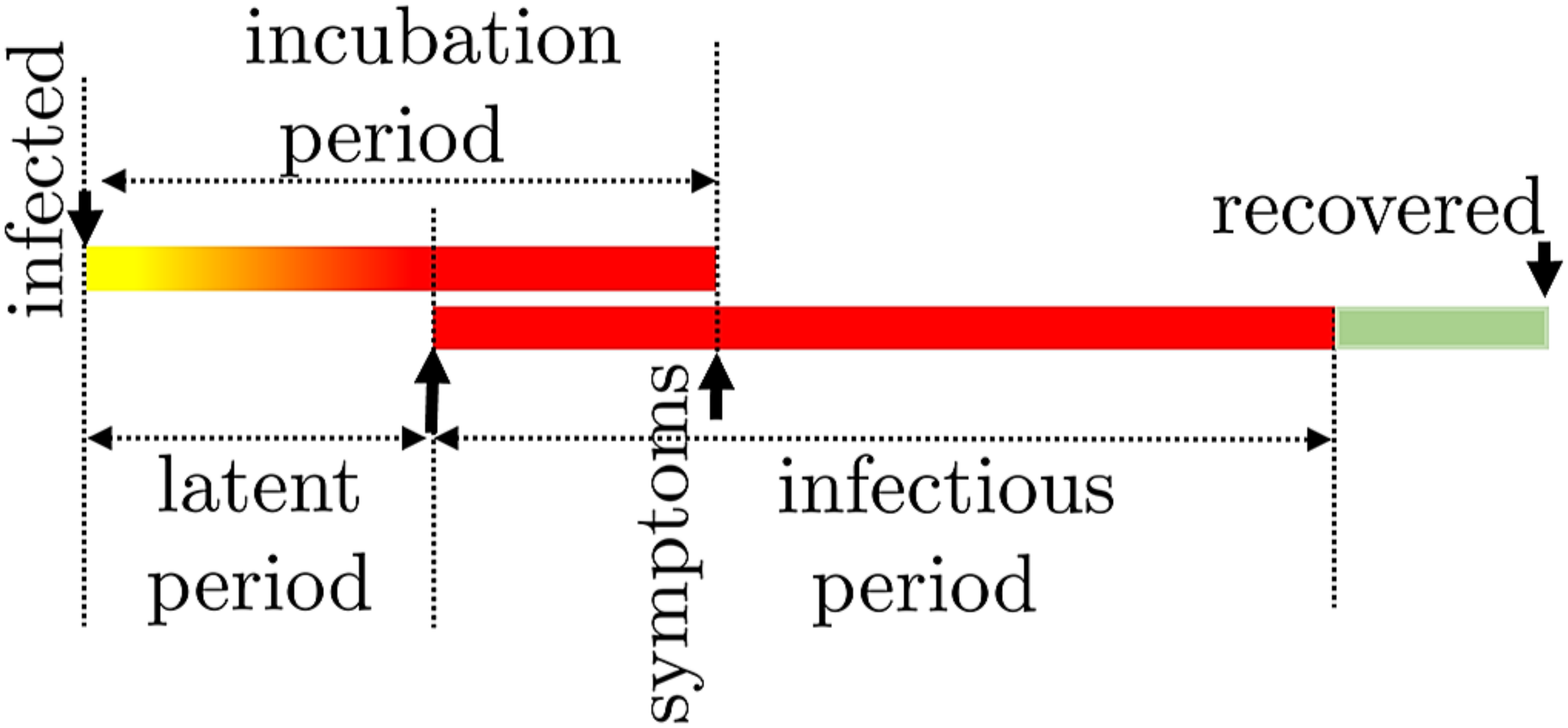}
    \caption{Infectious stages. Once contagious occurs, our interest centers on the duration of the incubation period, the infectious period, and the latent period.}
    \label{fig:stages}
\end{figure}

As a consequence of the Bayesian process, the posterior,$P(I_{t - \tau +1}, \dots, I_t, R_{t,\tau})$, has a mean value and standard deviation  for the Gamma distribution given by 
\begin{equation}
\mu = \frac{a + \sum_{s = t - \tau + 1}^{t} I_s}
{\frac{1}{b} + \sum_{s = t - \tau + 1}^{t} \Lambda_s},
\mbox{ and } 
\sigma =  \frac{a + \sum_{s = t - \tau + 1}^{t} I_s}
{\left(\frac{1}{b} + \sum_{s = t - \tau + 1}^{t} \Lambda_s\right)^2}.
\end{equation}

Once exposed, a person infected passes through a series of stages of interest for our modeling (see Figure~\ref{fig:stages}). Practitioners call {\it incubation period} to the time between the acquisition of the virus and the onset of symptoms. Also, they call  {\it latent period} to the time between the acquisition of the infectious and the beginning of the infectious period. Please note that if the latent period is less than the incubation period, the person infected may start infecting others without showing symptoms. Still, there seem to be people passing through the infectious period without showing symptoms~\cite{nishiura2020estimation}. Indeed it may be the case that people may start recovering before or after the infectious period, but that does not affect Cori {\it et al.}~\cite{cori2013new} model.

\begin{figure*}
    \centering
    \begin{tabular}{c}
\includegraphics[width=6in]{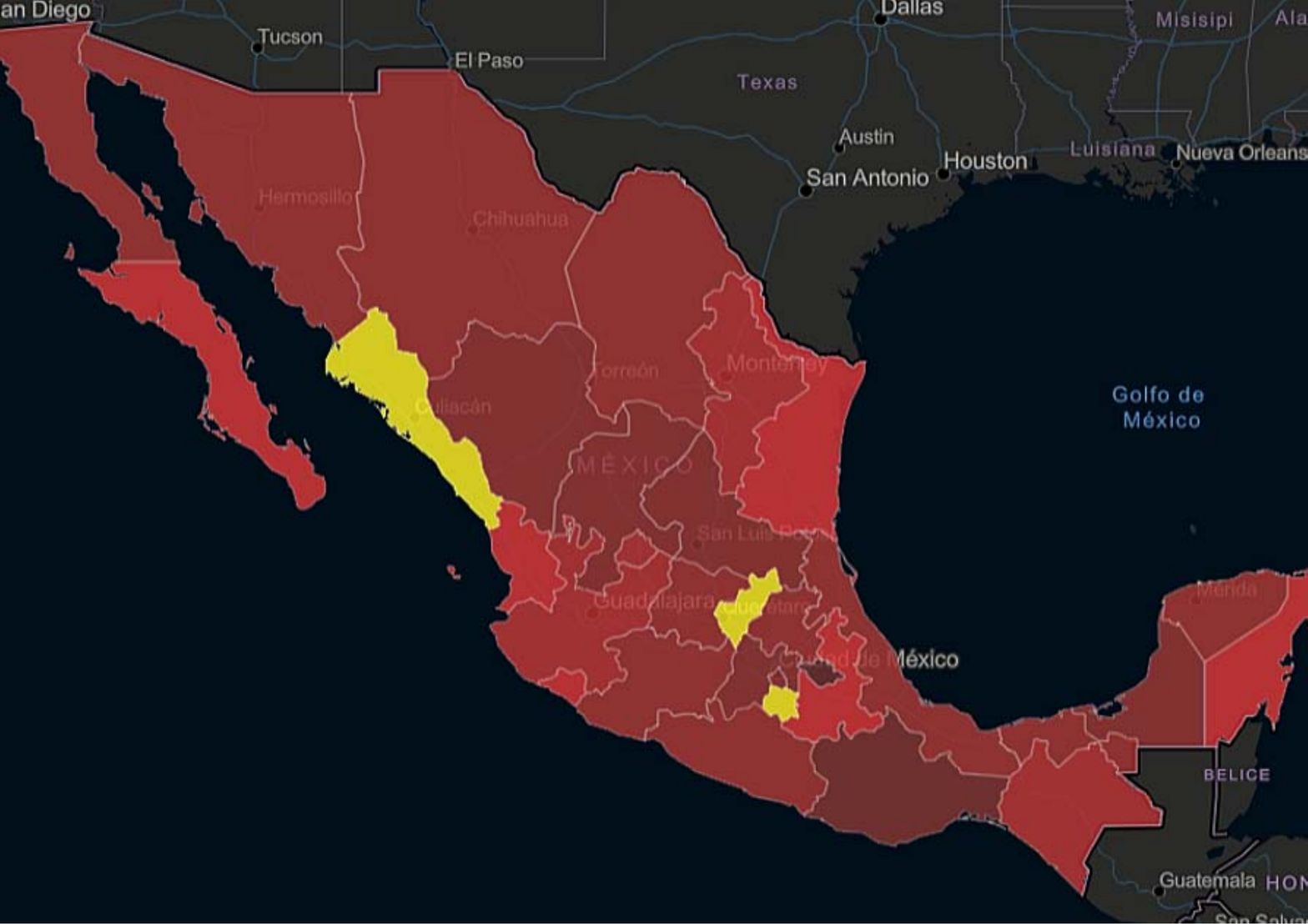} \\
(a) \\
\includegraphics[width=6in]{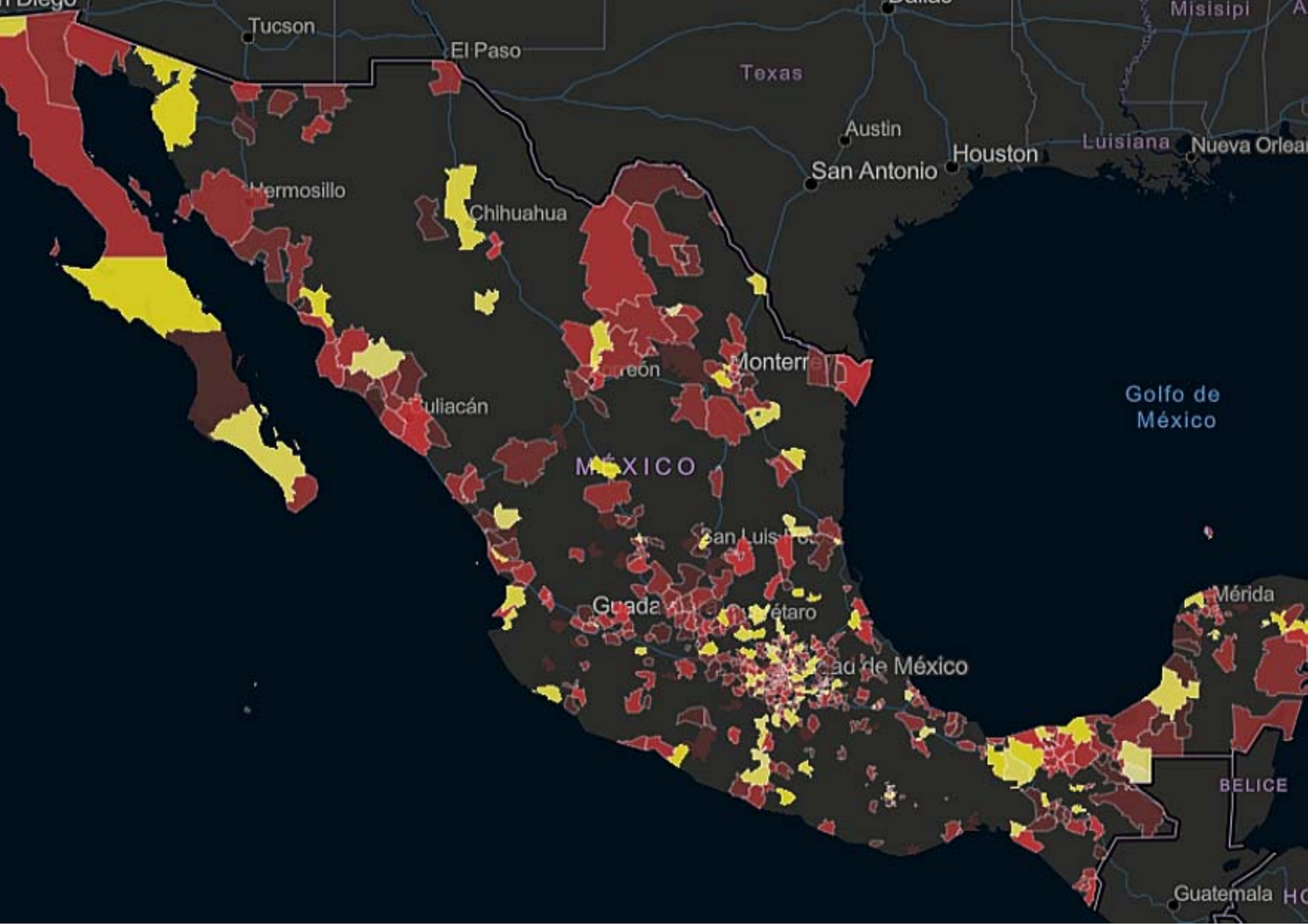}\\ 
 (b)\\
    \end{tabular}
    \caption{Basic Reproduction Number $R-t$  of COVID-19 in Mexico at state (a) and municipal  (b) level.  }
    \label{fig:covid-spread}
\end{figure*}

\section{Results}
We implemented Cori {\it et al.}~\cite{cori2013new} model using the {\bf R} (the language) library {\it EpiEstim}. Cori {\it et al.} assumed that the rate at which a person infects others follows a Poisson process. They then use a Bayesian inference process where they assume a {\it prior} in the form of a Gamma distribution, resulting in a {\it  posterior } that also follows a Gamma distribution. We provide the resulting mean and standard deviation in our maps as the most plausible element of the estimate's behavior and uncertainty. Following the recommendation of Cori {\it et al.}, we calculate $ R_t $ only when the number of cases is greater than 12.  Also, using the research by Aghaali {\it et al.}~\cite{aghaali2020estimation}, we assign a mean of 4.55 and a standard deviation of 3.3 to the  Gamma {\it prior}. We are noting that the characterization of the serial interval varies widely and possibly requires values that reflect the dynamics of the country~\cite{griffin2020rapid}.

\begin{figure}[h]
    \centering
    \includegraphics[width=3in]{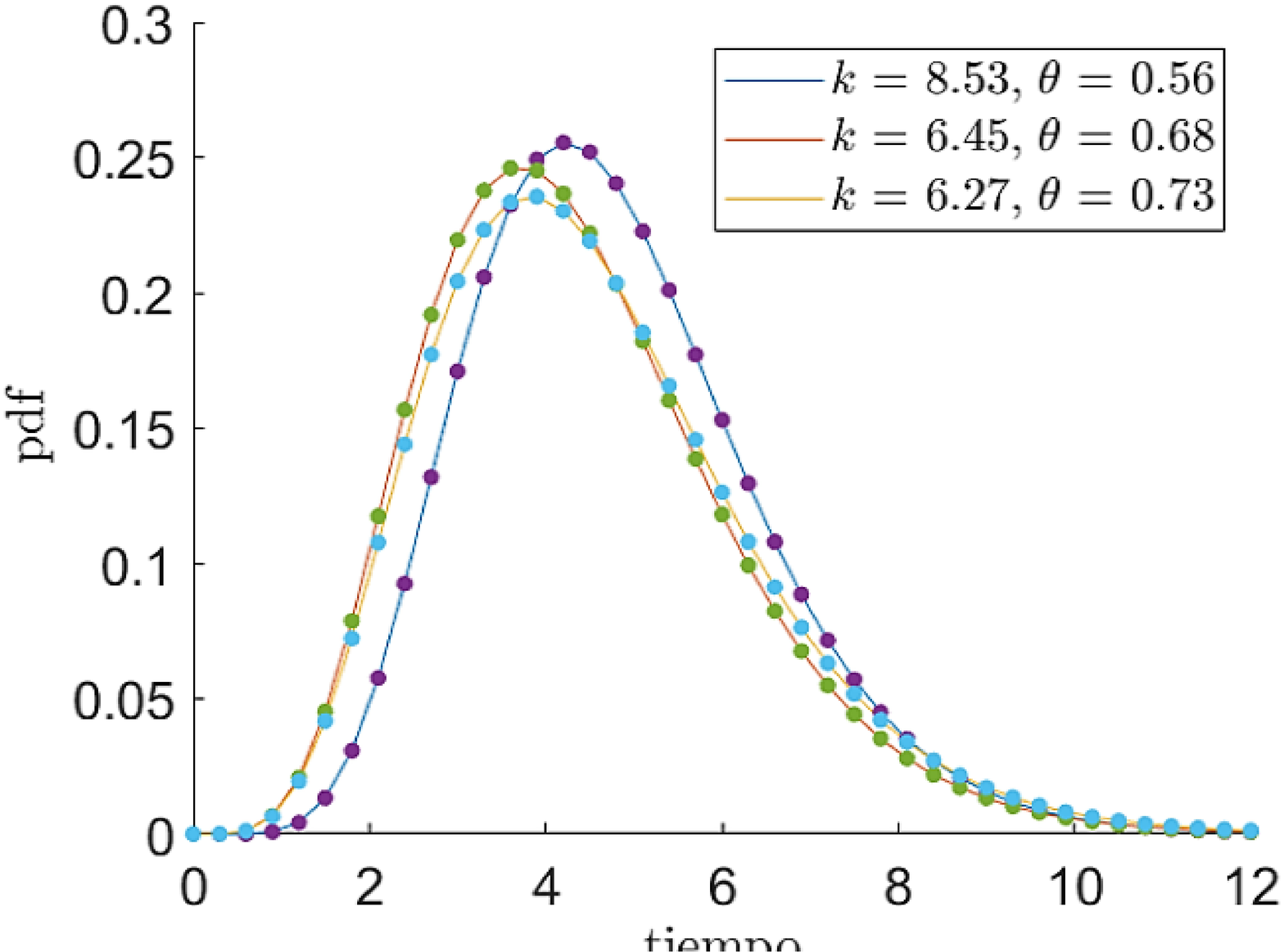}
    \caption{Gamma Prior. Illustration for the Gamma distribution for some parameters in the literature.}
    \label{fig:shapeScale}
\end{figure}

The model requires the definition of parameters, including the incubation, the latent, and the infectious periods. Once infected, people seem to have an incubation period with means that range from 5.1 to 5.84 days,  with 95\% of the population covered after a means ranging from 11.5 to 12.5 days~\cite{lauer2020incubation, han2020estimate, li2020early}. Still, there are reports of ultralong incubation periods of 38 days~\cite{wang2020case}, giving rise to the need for more studies~\cite{cimolai2020more}. There is still some debate on whether infected people start spreading COVID-19 before showing symptoms~\cite{bae2020chinese,yu2020familial}, although there may be some evidence tilting toward that conjecture with ranges between one and four days. Byrne {\it et al.}~\cite{byrne2020inferred} compiled 48 articles to define the period of infectious. Interestingly, they found that for asymptomatic people, the median ranges from 6.5 to 9.5 days; for not hospitalized, the median is 13.4 days, and for hospitalized people, it is 18.1 days. Thus, in our model, we use $\tau$ equal to 13.4 days.

One also needs to provide the parameters for the Gamma prior. There is a significant problem with obtaining these values, as one needs to be able to trace contacts for many people carrying an infectious. Nishiura {\it et al. } ~\cite{nishiura2020serial}, Zhao {\it et al.}~\cite{zhao2020serial}, and Aghaali {\it et al.}~\cite{aghaali2020estimation} calculated a mean value  $\mu$ between 4.4 and 4.8 and standard deviation $\sigma$ between  2.7 and 3.3, with a number of cases studied $n$ varying between  18 and 318 (see Figure ~\ref{fig:shapeScale}). Still, Griffin {\it et al.}~\cite{griffin2020rapid} found a variation for $\mu$ between 3.1 and 7.5 in 22 estimates.


\begin{figure}[t]
    \centering
    \includegraphics[width=4in]{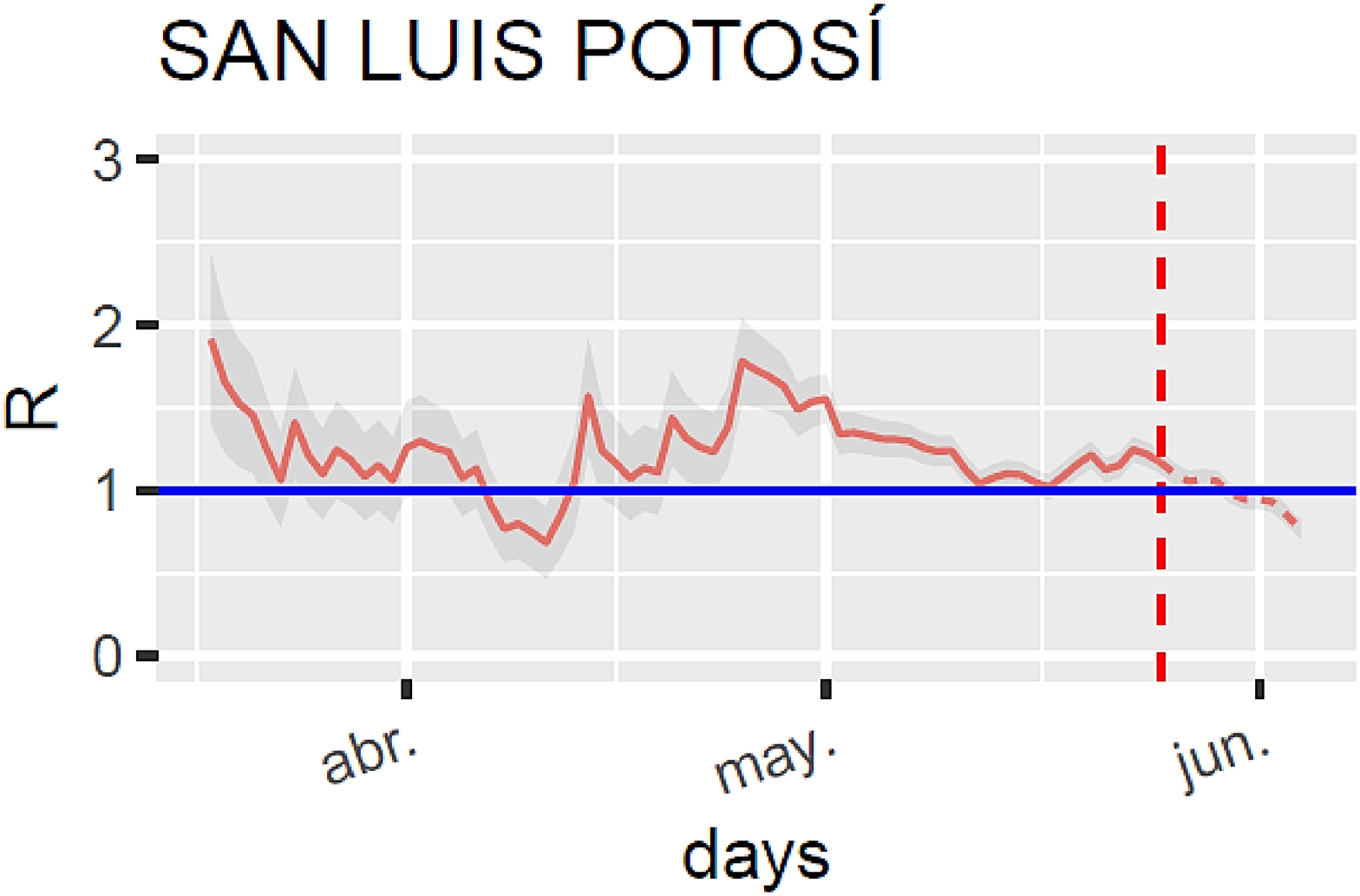}
    \caption{Daily Estimate of $R_t$. Here, we plot $R_t$ mean value and the 95\% CI under a shaded area around it. We plot a horizontal blue line for $R_t = 1$, and a vertical red dashed line at the furthest date in the past where there was a change between the current report the one presented one day before. From that day on, we plot with a dashed line the estimate of the mean value of $R_t$. }
    \label{fig:RtEstimate}
\end{figure}

We display $R_t$ as choropleth maps at municipal and state levels of granularity using a red shading pattern for polygons whose value is equal to or greater than one and a yellow to white shading pattern whose value is below one (see Figure~\ref{fig:covid-spread}).
To construct the charts, for each entity under analysis, we compared the records between the most current data set and the previous day. Then, we found the most remote date when there was a change from one day to the next in the number of confirmed cases. Finally, we show both a vertical dashed line on that date, and from that day on, we continue the estimation of $R_t$ with a dashed line for the rest of the timeline. The estimate of $ R_t $ has an uncertainty range that we shadow around the mean (see Figure~\ref{fig:RtEstimate}).

One can access our maps to visualize the number of confirmed, suspected, and deceases cases at \url{https://tinyurl.com/mexico-covid19-count}, while one access the map to visualize the estimation of $R_t$  at \url{https://tinyurl.com/mexico-Rt}. In both cases, we used ArcGIS as the software platform.

\section*{Conclusion}
There is an identified need to learn locally about the spread of COVID-19 to make informed decisions. Using official data publicly sourced by the Mexican Health Ministry, we introduced two visualizations: In one, we map the magnitude of positively confirmed, pending, and deceased cases and in another where we mapped the basic reproduction number $R_t$, both at state and municipal level. Combined, they provide complete information about the magnitude and direction of the epidemy. Daily, we update the maps as new information becomes available.

As our understanding of COVID-19 increases, there is the opportunity to incorporate that knowledge into visualization tools. Researchers and practitioners in areas as diverse as epidemiology, sociology, 
economy, and computing may bring new ideas to tackle the pandemic's consequences. As further information becomes available, dynamic visualization tools may provide a form of rapid form of communication to gain insight. This is especially true in the local information case as it may provide critical insights about the dynamics, culture, and phenomena impacting the spread of COVID-19.

\section*{Acknowledges}
Thanks to the Mexican Health Ministry for making publicly available the dataset on top of which we construct this work.

\appendix
 \section{Data Dictionary}
 \label{app:data}

 The features used in our analysis are the result of the collection of the samples reported to the National Network of Epidemiological Surveillance Laboratories. The data  include the following fields~\cite{dataset}:

 \begin{center}

 \begin{tabular}{cll}
 \rowcolor[gray]{0.5}
 {\bf ID} & {\bf Name} & {\bf Concept}\\
 1 & {\tt update.date} &
\begin{minipage}{4in}
Date of release of the dataset.
\end{minipage}\\ 
\rowcolor[gray]{0.9} 2 & {\tt record.number} &
\begin{minipage}{4in}
 Unique identifier of a person's record.
\end{minipage}\\ 
3 & {\tt origin} &
\begin{minipage}{4in}
Whether the medical record was captured through the Respiratory Disease Monitoring Health Unit System (USMER) or not.  
\end{minipage}\\  

 \rowcolor[gray]{0.9}
4 & 	
{\tt sector} &
\begin{minipage}{4in}
Institution of the National Health System providing care.
\end{minipage}\\                                                               
5 &
{\tt state.med.unit} &
\begin{minipage}{4in}
State code location for the medical unit.  
\end{minipage}\\
 \rowcolor[gray]{0.9}
 6 &
{\tt sex} &
\begin{minipage}{4in}
Genre of the patient.  
\end{minipage}\\         
7 & 
{\tt state.res} &
\begin{minipage}{4in}
Patient's residence State name.
\end{minipage}\\
\rowcolor[gray]{0.9} 8 & 
{\tt municipio.res} &
\begin{minipage}{4in}
Patient's Municipio of residence.
\end{minipage}\\

9&
{\tt patient.type} &
\begin{minipage}{4in}
Whether the  patient is outpatient or inpatient.
\end{minipage}\\

\rowcolor[gray]{0.9}10 & 
{\tt symptoms.date} &
\begin{minipage}{4in}
Date when the patient experienced the first symptoms.
\end{minipage}\\

11	&
{\tt decease.date} &
\begin{minipage}{4in}
Indicates the date of patient decease, if it applies. Otherwise, it remains empty if the patient is alive.
\end{minipage}
\\

\rowcolor[gray]{0.9}         
12	&
{\tt pneumonia} &
\begin{minipage}{4in}
Whether a physician diagnosed the patient with pneumonia. 
\end{minipage}\\                                                                 
13&
{\tt age} &
\begin{minipage}{4in}
Age of the patient in years.
\end{minipage}\\
\rowcolor[gray]{0.9}
14 &
{\tt pregnant} &
\begin{minipage}{4in}
Whether the patient is pregnant.                                                 \end{minipage}              \\

15 & 
{\tt diabetes} &
\begin{minipage}{4in}
Identifies whether a physician diagnosed the patient with diabetes. 
\end{minipage}                               \\                       \rowcolor[gray]{0.9}    
16 & 
{\tt COPD} &
\begin{minipage}{4in}
Does the  patient has  Chronic Obstructive Pulmonary Disease?
\end{minipage}                                        \\             
17	&
{\tt asthma} &
\begin{minipage}{4in}
Whether a physician diagnosed the patient with  asthma.
\end{minipage}\\                                                      \rowcolor[gray]{0.9}           
18	& 
{\tt immunosuppression} &
\begin{minipage}{4in}
Whether a physician diagnosed the patient with immunosuppression.
\end{minipage}\\                                                               
19 & 
{\tt hypertension} &
\begin{minipage}{4in}
Whether a physician diagnosed the patient with hypertension. 
\end{minipage}\\                                                       \rowcolor[gray]{0.9}          
20 & 
{\tt other.diseases} &
\begin{minipage}{4in}
Whether a physician diagnosed the patient   with other diseases.  
\end{minipage}\\                                                                 
21 & 
{\tt cardiovascular} &
\begin{minipage}{4in}
Whether a physician diagnosed the patient with a cardiovascular disease.
\end{minipage}\\                                                      \rowcolor[gray]{0.9}            
22 & 
{\tt obesity} &
\begin{minipage}{4in}
Identifies whether a physician diagnosed the patient   with a probem of obesity.
\end{minipage}\\                                                                 
23	 & 
{\tt chronic.kidney} &
\begin{minipage}{4in}
Whether a physician diagnosed the patient   with chronic kidney failure.
\end{minipage}\\                                                      \rowcolor[gray]{0.9}           
24 & 	
{\tt smoking} &
\begin{minipage}{4in}
Whether a physician diagnosed the patient with a smoking habit.
\end{minipage}\\                                                                  
25	&
{\tt contact} &
\begin{minipage}{4in}
Whether    the patient  made contact  with another case diagnosed with SARS CoV-2.
\end{minipage}\\                   \rowcolor[gray]{0.9} 
\rowcolor[gray]{0.9}
26	&
{\tt result} &
\begin{minipage}{4in}
Whether the patient has a positive, negative or pending test for SARS-CoV-2.
\end{minipage}
\\


\end{tabular}
\end{center}

\bibliographystyle{unsrt}
\bibliography{references}

\end{document}